\documentclass[3p]{elsarticle}

\usepackage{amssymb}
\usepackage{amsmath}
\usepackage{graphicx}
\usepackage{ucs}
\usepackage{url}
\usepackage[utf8x]{inputenc}
\usepackage{subfigure}
\usepackage{booktabs}

\journal{Physica A}

\begin{document}

\begin{frontmatter}



\title{Stock-flow consistent macroeconomic model with nonuniform distributional constraint}

\author[label1,label2]{Aurélien Hazan}
\address[label1]{Université Paris-Est}
\address[label2]{LISSI, UPEC, 94400 Vitry sur Seine, France}



\begin{abstract}
We report on results concerning  a partially aggregated Stock Flow Consistent (SFC)
macro-economic model in the stationary state where the sectors of banks and
firms are aggregated, the sector of households is disaggregated, and the probability
density function (pdf) of the wealth of households is exogenous, constrained by econometric data.
It is shown that the equality part of the constraint can be reduced to a single
constant-sum equation, which relates this problem to the study of continuous 
mass transport problems, and to the sum of iid random variables. 
Existing results can thus be applied, and provide marginal probabilities, and the
location of the critical point before condensation occurs.
Various numerical experiments are performed using Monte Carlo
sampling of the hit-and-run type, using wealth and income data for France.
\end{abstract}


\end{frontmatter}




While the neoclassical standpoint on macroeconomics has adapted in response to growing criticism 
in the aftermath of the subprime crisis \cite{benes_financial_2014}, it still relies on 
intensely debated hypotheses, for example rational expectations, utility 
functions, or representative agents, that form the building blocks of Dynamic Stochastic
General Equilibrium (DSGE).

Other modelling approaches include ``the Agent Based (\ldots) approach,
which conceives the economy as a complex adaptive system populated by heterogeneous locally
interacting agents, and the Stock Flow Consistent framework (\ldots), which provides a 
comprehensive and fully integrated representation of the real and financial sides of the
economy through the adoption of rigorous accounting rules based on the quadruple entry
principle developped by Copeland'' \cite{caiani_agent_2016}.

SFC macroeconomic models, that date back to the 1950's, enforce local conservation
of money in a holistic perspective. They usually stand at the aggregate level, and cover a
large scope of economic phenomena \cite{caverzasi_post-keynesian_2015}.
Thanks to quadruple-entry bookkeeping, they are able to guarantee that real
and financial transactions, involving many agents belonging to different sectors,
stay balanced \cite[2.6]{godley_monetary_2007}.

Agent Based Models (ABM), on the other hand, adopt a bottom-up perspective, and are able to
recover stylized facts, starting from low-level discrete-time dynamical description.
Most of the time, closed form solutions are not available for ABM 
(though recent works anchored in statistical physics started to fill that gap
\cite{gualdi_tipping_2015,carvalho_income_2014}). 

The link between SFC and microeconomic models started to be explored during the 1970's
\cite{seppecher_modeles_2016} 
and is now an active research topic. Agent Based Models that enforce SFC rules (SFC-AB)
have been studied over the last years in the post-keynesian community.
They rely on economic models, multi-agent programming, and a critical phase of calibration
\cite{caiani_agent_2016}. Little theoretical results exist to explain the behavior of such systems.

In this article, we look for a tradeoff between the complexity of the model, and the availability 
of theoretical results. Our interest goes to increasing the number of agents, with linear
behaviors, in a stationary state. What can be said at the population level, that is concerning
the distribution of the different variables ?

To answer this question, we first look at theoretical properties available for SFC models: 
mathematically, SFC can be formalized as a system of difference equations. Depending on the nature
of transactions flows, an analytic expression can be found for the stationary state (if any) and
for time-dependent transients \cite{godley_monetary_2007} of aggregated models. 
In \cite{hazan_volume_2017}, the stationary state of a particular,
fully-disaggregated SFC model with random connectivity, was seen as a Constraint Satisfaction Problem
(CSP), with solutions lying inside a random convex polytope. Marginal pdf of money stocks and flows
were numerically estimated, and the effect of various flow knockouts was evaluated
in the case of a BMW model \cite[7]{godley_monetary_2007} which describes an economy where
private banks create money through loans.

In the present article another stand is taken: while the accounting constraints are maintained,
the marginal pdf of several variables is considered \emph{exogenous}, that is constrained 
so as to respect empirically observed data. The marginal pdf of the other variables still is 
an unknown that we seek to approximate. This approach differs from recent SFC-AB works 
e.g. \cite{caiani_agent_2016} where an initial uniform distribution is supposed.

We thus report on the following original results: 
in the case of a partially aggregated SFC model of the BMW type where banks and firms are 
represented by one agent respectively, the set of accounting identities that define the SFC model
can be reduced to a constant sum over individual wealth $M_i$, the distribution of which is 
constrained by a weight function $f(m)$.
The properties of the sum $\sum M_i$ are discussed, using previous investigations
by \cite{evans_canonical_2006, filiasi_concentration_2014,burda_wealth_2002} that concluded 
to the existence of a phase transition leading to condensation. The marginal single site pdf
of all variables are deduced, for various usual weight functions. 
Furthermore a simple Monte Carlo sampling algorithm is available, and allows to simulate the
behavior of a simple economy.

Section \ref{theory} depicts the SFC model, the distributional constraints imposed, and the simplified
constant sum problem obtained after gaussian elimination ; section \ref{sec.mass.transport} discusses
how the latter is related to continuous mass transport studies in the statistical physics litterature.
Numerical simulations and results are examined in section \ref{sec.numeric.experiments}. 
Section \ref{sec.discussion} and \ref{sec.conclusion} conclude.

\section{Partially aggregated SFC model with constrained pdf}
\label{theory}

In this section, the elements of the BMW model are briefly exposed: the balance sheet, the
transaction matrix that define the flows of money occurring between an origin and a destination,
and the behavioral equations that specify the flows. Notations follow Godley and Lavoie \cite{godley_monetary_2007} and
are explained in Tab. \ref{tab.bmw short labels}, more detail concerning accounting conventions can be found in Appendix.
Tab. \ref{tab.balance} represents a balance sheet, that is the summary of assets, liabilities, and capital of all
agents, at a fixed time such as the end of the year. The particular model is a partially aggregated
BMW model, with private money, no state nor central bank, where agents are grouped by sector (firms,
banks, households). More detail can be found in \cite[7]{godley_monetary_2007}, and a fully disaggregated version is discussed
in \cite{hazan_volume_2017}. It follows from this simplified setup that the stock of capital $K$, the stock of loans $L$,
and the sum of deposits $\sum M_i$ are equal.

\begin{table}[htbp]
	\centering
	\begin{tabular}{lllllll} %
	\hline
	& \multicolumn{3}{c}{Households}  & \multicolumn{1}{c}{Firms} & \multicolumn{1}{c}{Banks} & $\sum$ \\
	\cmidrule(lr{.75em}){2-4}  \cmidrule(lr{.75em}){5-5}  \cmidrule(lr{.75em}){6-6}
	 & 1 & 2 & 3 & 1 & 1  &\\
	\hline
	Money deposits & $M_1$  &  &  &    & -$M_1$ & 0\\
		           &        & $M_2$    &  & &-$M_2$  & 0\\	
		           &        &  & $M_3$   &  & -$M_3$ &  0 \\	
	\hline
	Loans  &   &  &  & -$L_1$ &  $L_1$  & 0\\		           
	\hline
	Fixed capital  &   &  &  & $K_1$ &   & $K_1$\\		           
	\hline               
	Balance (net worth)  & -$V_{h1}$  & -$V_{h2}$ & -$V_{h3}$ &   &  & -$\sum_i V_{hi}$ \\		                          		
	\hline
	$\sum$               & 0  &0  &0  & 0 & 0 & 0\\		        
	\hline
	\end{tabular}
\caption{\label{tab.balance} Example of balance sheet of the BMW model with many households, one bank and one firm: $nw=3$, $nf=1$, $nb=1$.
$M_i,L_j,K_k$ are the individual money deposits, loans, and tangible capital.
$V_{hi}$ is the net worth of individual households.}
\end{table}

During the period (say a year) that separates two balance sheets, many transactions occur such as
consumption, investment, wage, depreciation, interest (on loans and deposits). The transactions
corresponding to the partially aggregated BMW model are summarized in Tab. \ref{bmw.transactions},
and explained in a detailed way in the Appendix. The balance sheet at time $t+1$ is the result of applying these
transactions to the balance sheet at time t.

Finally, the behavioral equations are :
\begin{eqnarray}
AF &=& \delta  K_{t-1} \label{eq.AF} \\
C_{d,i} &=& \alpha_{0,i} + \alpha_1  YD_{i} + \alpha_2  M_{t-1,i} \label{eq.Cd} \\
I_d &=& \gamma (\kappa Y_{t-1} - K_{t-1} ) + AF \label{eq.Id}
\end{eqnarray}

Eq.(\ref{eq.AF}) states that the depreciation of tangible capital is proportional to its past stock $K_{-1}$,
eq.(\ref{eq.Cd}) is the consumption demand $C_{d,i}$ of household $i \in [1,nw]$ which is a mixture of an 
autonomous term, of the disposable income $YD_i$ and of the wealth accumulated by households $M_{t-1,i}$
at the previous time step. Lastly, eq.(\ref{eq.Id}) sets a target investment level, that depends both on
amortization and on some target capital $kY_{t-1}$, where $Y_{t}$ is the  is the total production at
time $t$, defined as the sum of consumption supply $Cs$ and investment supply $Is$. 

In the steady-state regime considered in this article, the term $\kappa Y_{t-1} - K_{t-1}$ 
vanishes as well as changes in deposits and loans :

\begin{eqnarray}
\Delta M_i &=& 0 \\
\Delta L &=& 0 \\
I_d &=&  AF =\delta  K_{−1}
\end{eqnarray}

Notations are gathered in Table \ref{tab.bmw short labels}. The subscripts $d$ and $s$ stand for
demand and supply, $i$ represents quantities related to individual households, $t$ stands for
the time instant.

\begin{table}[htbp]
\centering
\begin{tabular}{ll}
\hline
 Variable & Label\\
\hline
 money deposit & $M$ \\
 capital & $K$ \\
 loans to firms  & $L$   \\
 investment & $I$  \\
 interest on loans & $IL$  \\
 wage bill & $WB$   \\
 depreciation allowance & $AF$  \\
 interest on workers deposits & $ID$  \\ 
 consumption of workers & $C$  \\
\end{tabular}
\caption{Labels associated with the different monetary variables, after 
\cite{godley_monetary_2007}. The subscripts $d$ and $s$ stand for demand and supply, $i$ represents quantities related to individual households, $t$ stands for the time instant.}
\label{tab.bmw short labels}
\end{table}

\subsection{Properties of the partially aggregated model}
\label{sub.properties}

In the steady-state unidimensional case, Godley and Lavoie find that $\sum M =k Y$ and
$Y=\frac{\alpha_0}{(1-\alpha_1)(1-\delta k) -k \alpha_2} $.

Let $A'$ be the augmented matrix of the system $Ax=b$. 
In the case $nw=2$, with $\alpha_{0,1}=\alpha_{0,2}=\alpha_{0}$:
\begin{equation}           	
A'=
\begin{bmatrix}
1 & 0 & 0 & 0 & 0 & 0 & 0 & -1 & -1
& 0 & 0 & 0 & 0 & 0 \\
0 & 0 & 0 & 1 & -1 & 0 & 0 & 0 & 0 &
0 & 0 & 0 & 0 & 0 \\
-1 & 0 & r & -1 & 0 & 1 & 0 & 0 & 0
& 1 & 1 & 0 & 0 & 0 \\
0 & - \delta & 0 & 0 & 0 & 1 & 0 & 0 & 0
& 0 & 0 & 0 & 0 & 0 \\
0 & 0 & 0 & 0 & 1 & -1 & 0 & 0 & 0 &
0 & 0 & 0 & 0 & 0 \\
0 & 0 & 0 & 0 & 0 & 0 & 0 & -1 & 0 &
1 & 0 & r & 0 & 0 \\
0 & 0 & 0 & 0 & 0 & 0 & 0 & 0 & -1 &
0 & 1 & 0 & r & 0 \\
0 & 0 & 0 & 0 & 0 & 0 & 0 & -1 & 0 &
\alpha_1 & 0 & r \alpha_1 + \alpha_2 &
0 & \alpha_0 \\
0 & 0 & 0 & 0 & 0 & 0 & 0 & 0 & -1 &
0 & \alpha_1 & 0 & r \alpha_1 +
\alpha_2 & \alpha_0 \\
0 & 0 & 0 & 0 & 1 & 0 & -1 & 0 & 0 &
0 & 0 & 0 & 0 & 0 \\
0 & - \delta & 0 & 0 & 0 & 0 & 1 & 0 & 0
& 0 & 0 & 0 & 0 & 0 \\
- k & 1 & 0 & - k & 0 & 0 & 0 & 0 & 0
& 0 & 0 & 0 & 0 & 0 \\
0 & 1 & -1 & 0 & 0 & 0 & 0 & 0 & 0 &
0 & 0 & 0 & 0 & 0 \\
0 & 1 & 0 & 0 & 0 & 0 & 0 & 0 & 0 &
0 & 0 & -1 & -1 & 0
\end{bmatrix}
\end{equation}

The reduced row echelon form (rref) of $A'$ is:      	
\begin{equation}           
A'_{\textrm{rref}}=
\left(\begin{array}{rrrrrrrrrrrrrr}
1 & 0 & 0 & 0 & 0 & 0 & 0 & 0 & 0 &
0 & 0 & 0 & 0 & \frac{-2 \delta k \alpha_0 + 2
\alpha_0}{- \delta k \alpha_1 + \delta k + k
\alpha_2 + \alpha_1 - 1} \\
0 & 1 & 0 & 0 & 0 & 0 & 0 & 0 & 0 &
0 & 0 & 0 & 0 & \frac{2 k \alpha_0}{- \delta k
\alpha_1 + \delta k + k \alpha_2 + \alpha_1 - 1} \\
0 & 0 & 1 & 0 & 0 & 0 & 0 & 0 & 0 &
0 & 0 & 0 & 0 & \frac{-2 k \alpha_0}{\delta k
\alpha_1 -  \delta k -  k \alpha_2 -  \alpha_1 + 1}
\\
0 & 0 & 0 & 1 & 0 & 0 & 0 & 0 & 0 &
0 & 0 & 0 & 0 & \frac{2 \delta k \alpha_0}{-
\delta k \alpha_1 + \delta k + k \alpha_2 +
\alpha_1 - 1} \\
0 & 0 & 0 & 0 & 1 & 0 & 0 & 0 & 0 &
0 & 0 & 0 & 0 & \frac{2 \delta k \alpha_0}{-
\delta k \alpha_1 + \delta k + k \alpha_2 +
\alpha_1 - 1} \\
0 & 0 & 0 & 0 & 0 & 1 & 0 & 0 & 0 &
0 & 0 & 0 & 0 & \frac{2 \delta k \alpha_0}{-
\delta k \alpha_1 + \delta k + k \alpha_2 +
\alpha_1 - 1} \\
0 & 0 & 0 & 0 & 0 & 0 & 1 & 0 & 0 &
0 & 0 & 0 & 0 & \frac{2 \delta k \alpha_0}{-
\delta k \alpha_1 + \delta k + k \alpha_2 +
\alpha_1 - 1} \\
0 & 0 & 0 & 0 & 0 & 0 & 0 & 1 & 0 &
0 & 0 & 0 & \frac{\alpha_2}{- \alpha_1 + 1}
& \frac{\delta k \alpha_0 \alpha_1 -  \delta k
\alpha_0 + k \alpha_0 \alpha_2 -  \alpha_0
\alpha_1 + \alpha_0}{\delta k \alpha_1^{2} - 2
\delta k \alpha_1 -  k \alpha_1 \alpha_2 + \delta k
-  \alpha_1^{2} + k \alpha_2 + 2 \alpha_1 - 1} \\
0 & 0 & 0 & 0 & 0 & 0 & 0 & 0 & 1 &
0 & 0 & 0 & \frac{- \alpha_2}{- \alpha_1 + 1}
& \frac{- \alpha_0}{- \alpha_1 + 1} \\
0 & 0 & 0 & 0 & 0 & 0 & 0 & 0 & 0 &
1 & 0 & 0 & \frac{r \alpha_1 -  r +
\alpha_2}{- \alpha_1 + 1} & \frac{2 r k
\alpha_0 \alpha_1 + \delta k \alpha_0
\alpha_1 - 2 r k \alpha_0 -  \delta k \alpha_0 + k
\alpha_0 \alpha_2 -  \alpha_0 \alpha_1 +
\alpha_0}{\delta k \alpha_1^{2} - 2 \delta k
\alpha_1 -  k \alpha_1 \alpha_2 + \delta k - 
\alpha_1^{2} + k \alpha_2 + 2 \alpha_1 - 1} \\
0 & 0 & 0 & 0 & 0 & 0 & 0 & 0 & 0 &
0 & 1 & 0 & \frac{- r \alpha_1 + r - 
\alpha_2}{- \alpha_1 + 1} & \frac{- \alpha_0}{-
\alpha_1 + 1} \\
0 & 0 & 0 & 0 & 0 & 0 & 0 & 0 & 0 &
0 & 0 & 1 & 1 & \frac{-2 k \alpha_0}{\delta k
\alpha_1 -  \delta k -  k \alpha_2 -  \alpha_1 + 1}
\\
0 & 0 & 0 & 0 & 0 & 0 & 0 & 0 & 0 &
0 & 0 & 0 & 0 & 0 \\
0 & 0 & 0 & 0 & 0 & 0 & 0 & 0 & 0 &
0 & 0 & 0 & 0 & 0
\end{array}\right)
\label{eq.rref}
\end{equation}

The last non-trivial row $M_1+M_2 = \frac{2k \alpha_0}{(1-\alpha_1)(1-\delta k) -k \alpha_2}$
is a constant sum equation that generalizes the unidimensional case, and has one degree of freedom.
Using the rref, all unknown variables can be computed when $M_1$ and $M_2$ are known.
In the general case $nw>2$, gaussian elimination gives the following result\footnote{this can be 
checked using a computer algebra system, see \url{https://gitlab.com/hazaa/sfc_proba}} :
\begin{eqnarray}
\sum_i^N M_i &=& \frac{k N \alpha_0}{(1-\alpha_1)(1-\delta k) -k \alpha_2} 
\label{eq.sum_Mi}
\end{eqnarray}

The dimension of the solution space of eq.(\ref{eq.sum_Mi}) is $nw-1$.
This relation holds in the particular case $nb=nf=1$, but is unlikely to remain valid when the
model is completely disaggregated, that is when $nf$ and $nb$ are greater than one.

Finally, from the system of equations in eq.(\ref{eq.rref}), one can get the following 
relation, which will be useful below, between wealth and income for a given agent:
\begin{eqnarray}
(1-\alpha_1) WB_{s,i} + (r (1-\alpha_1) -\alpha_2) M_i = \alpha_{0,i}
\label{eq.WBs_to_M}
\end{eqnarray}
\subsection{Modelling the distribution of wealth and income}
\label{sub.wealth.income}

The distribution of wealth and income display several regularities, as many authors
have shown since the 19th century. Economists such as V.~Pareto have studied the distribution
of wealth, and found a good fit to a power law among the richest.
Statisticians, and more recently, physicists became interested in the topic.
It it now well accepted that the bulk of the distribution of wealth and income fits a Gamma or
log-normal law. The findings of Pareto concerning the tail of the distribution, were
confirmed and refined by many subsequent studies. 
For incomes, various empirical estimations found a paretian tail $x^{-\gamma}$  with an exponent
$\gamma \in [1.5, 4]$ across time and countries \cite[p.25]{chakrabarti_econophysics_2013}.

In the case of France, income and wealth data are not available publicly
at the level of individuals or households. Recent empirical works by
economists and statisticians take advantage of tax tabulations to estimate the
percentiles of the distribution of wealth among the population \cite{fournier_generalized_2015}. 

\begin{figure}[htbp]
\centering
\subfigure[]{\includegraphics[width=6cm]{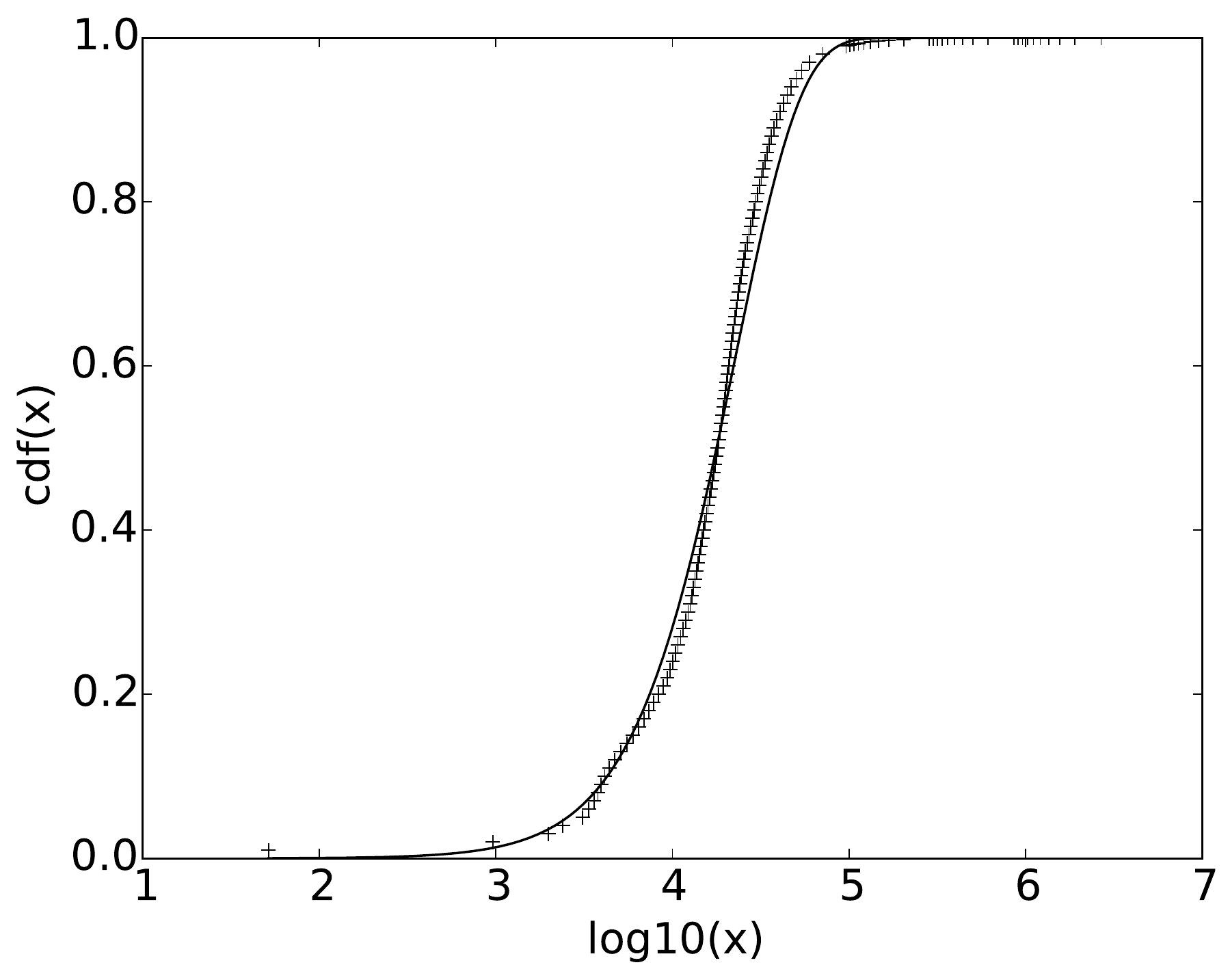}}
\subfigure[]{\includegraphics[width=6cm]{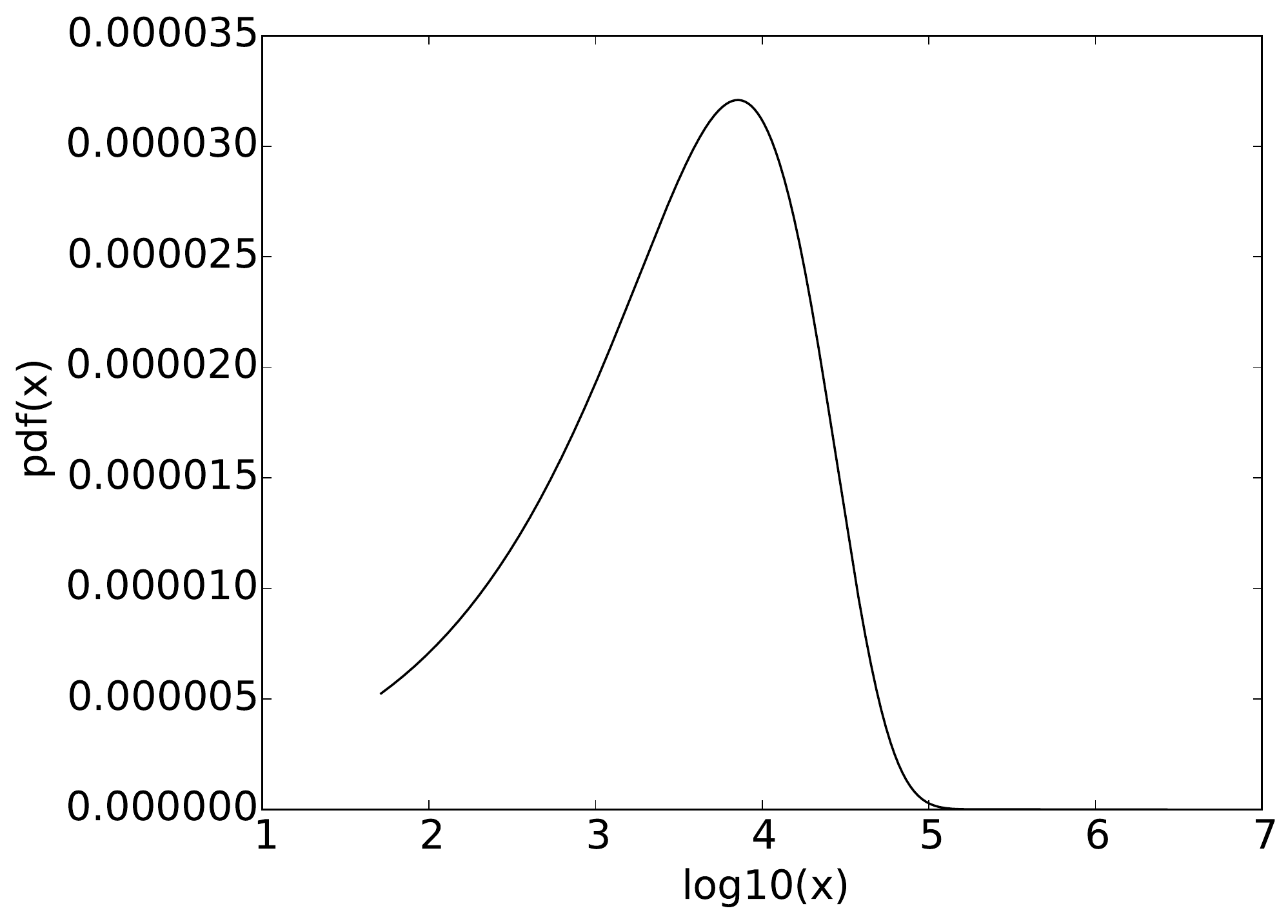}}
\subfigure[]{\includegraphics[width=6cm]{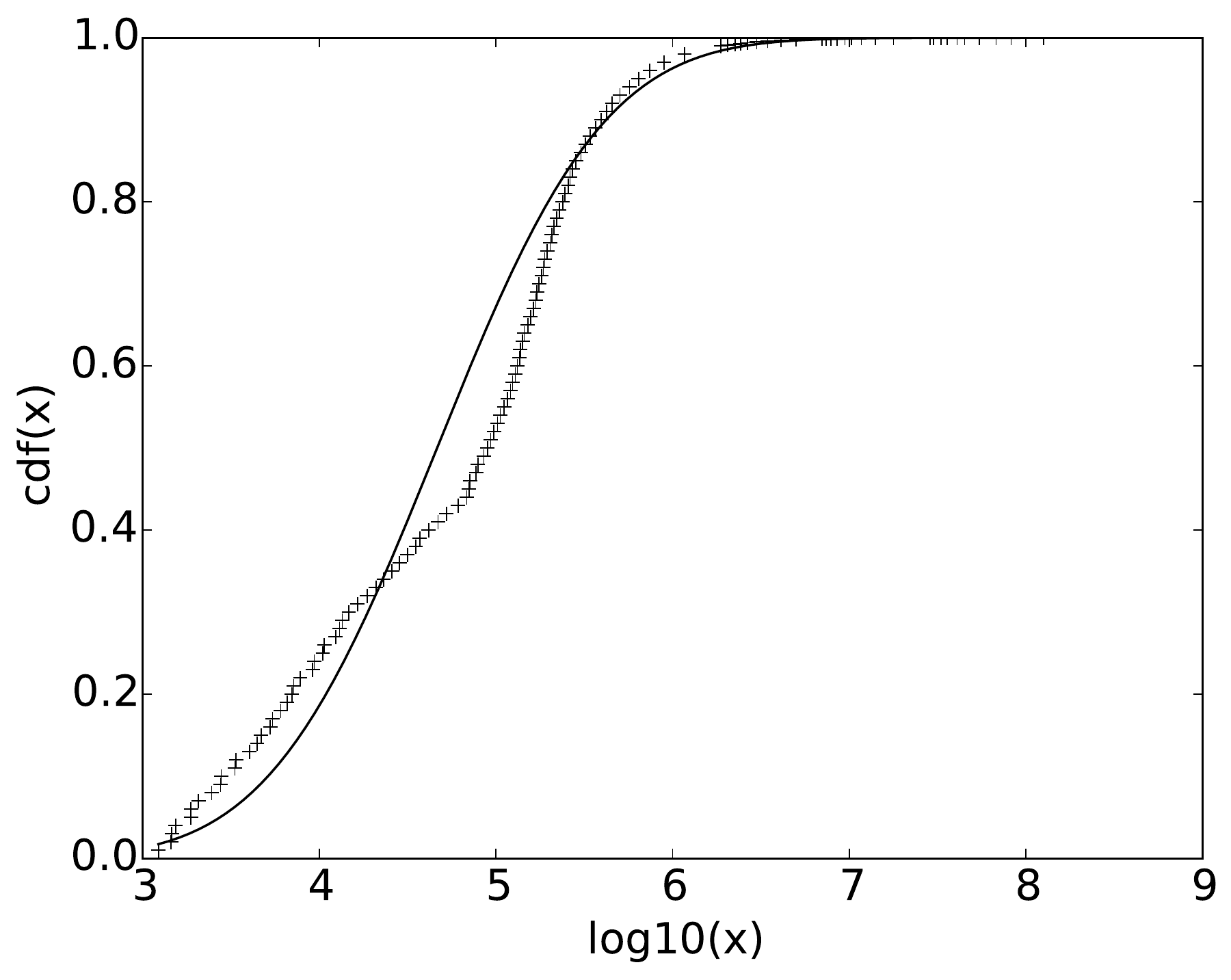}}
\subfigure[]{\includegraphics[width=6cm]{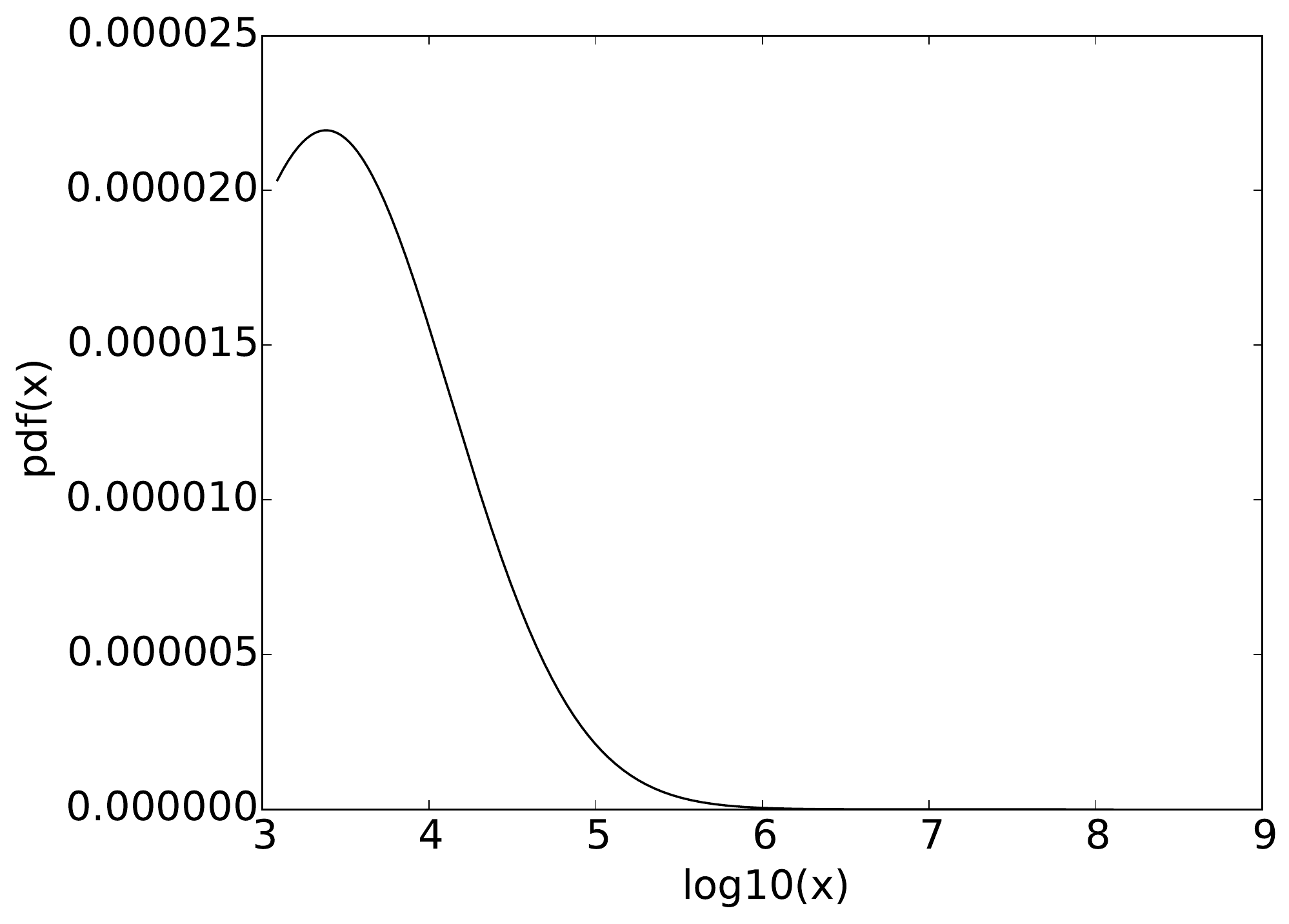}}
\caption{ (a) empirical cdf of fiscal annual income in France in the year 2010
represented by +, and in plain line the cdf of a Gamma variable with shape
parameter $a=1.46$ and scale parameter $1/\lambda=1.55 ~10^4$;
(b) pdf of the Gamma distribution;
(c) empirical cdf of wealth in France in the year 2010
represented by +, and in plain line the cdf of a lognormal variable with shape
parameter $a=1.72$ and scale parameter $4.64 ~10^4$;
(d) pdf of the lognormal distribution. }
\label{fig.distrib}
\end{figure}

Fig. \ref{fig.distrib}(a) shows the empirical cumulative distribution function (cdf)
of french income in 2010 provided by the World Inequality Database 
(see \ref{appendix.data}), as well as a gamma distribution with adequate parameters. 
Fig. \ref{fig.distrib}(a) shows the cdf of french wealth the same year, and a lognormal
fit. Fig. \ref{fig.distrib}(b,d) show the corresponding probability density functions (pdf).

Many nonparametric and parametric models of income and wealth have been discussed in 
the litterature \cite{chakrabarti_econophysics_2013}.
We compared several among them: a piecewise constant bulk with a Pareto tail \cite{feenberg_income_1992},
 a nonparametric method \cite{fournier_generalized_2015},
a monotone polynomial interpolation in the log domain \cite{murray_fast_2016},
and parametric models as shown in Fig \ref{fig.distrib}.
For simplicity, the distribution of income is modelled by a single Gamma distribution
on the full range covered by empirical data, and wealth by a lognormal distribution.
The fit of the income model to the empirical cdf is rather good, and sufficient 
for our needs in this article in the case of wealth. 

In sec. \ref{sec.numeric.experiments}, the topic of drawing random samples from a multivariate distribution,
with identically distributed components, given additional linear constraints, is discussed. 

\section{Analogy with continuous mass transport and sum or random variables models}
\label{sec.mass.transport}

Since some of the results obtained for mass transport models may be of interest to the
problem in sec. \ref{sub.properties}, let us summarize a few of them. 
The total mass, replaced by money, is constant and the number of sites is replaced by
the number of economic agents.
As pointed out in \cite{evans_canonical_2006, filiasi_concentration_2014} the problems
of mass transport and sum of random iid variables in the large deviation regime can be mapped.

Dynamical models of mass transport on a lattice have been the subject of many works:
it was observed that when the mass density increases above some critical threshold,
a condensed steady-state could appear under some distributional conditions.
Condensation means that a finite fraction of the total mass can concentrate on a single site.
The analogy between mass transport and macroeconomic modelling was noticed already
in \cite{burda_wealth_2002}.

To explain this phenomenon, it was first remarked that for some dynamical models (such as the ZRP), 
a factorized steady-state exists and can be computed. 
In that case, examined in \cite{bialas_condensation_1997,evans_canonical_2006}, the steady-state 
probability and the partition function write:
\begin{eqnarray}
P(\{m_l\}) &=& Z(M,L)^{-1} \prod_{l=1}^{L} f(m_l) \delta \Big( \sum_{l=1}^{L} m_l -M \Big) \\
Z(M,L) &=&   \prod_{l=1}^{L} \Big[ \int_0^{\infty} dm_l f(m_l) \Big]  \delta \Big( \sum_{l=1}^{L} m_l -M \Big)
\label{eq.Z}
\end{eqnarray}
where $f(m_l)$ is the single-site weight, $M$ is the total mass, $L$ the number of sites,
and $\delta \Big( \sum_{l=1}^{L} m_l -M \Big)$ embodies the constant sum constraint. 
When $f(m_l)$ is normalised, $Z(M,L)$ in eq.(\ref{eq.Z}) is the probability that a sum of $L$ random variables with pdf $f(m_l)$ sum to $M$.
The single-site weight $f(m_l)$ can be related to the distribution of wealth or income 
in sec. \ref{sub.wealth.income}, and to $p(m)$, the single-site marginal probability.
$p(m)$ is an unknown, that can differ from $f(m)$, because it takes into account the
fixed total amount of money:
\begin{eqnarray} 
p(m) &=& \int dm_2\ldots dm_L P(m,m_2,\ldots m_L) \delta \Big( \sum_{l=2}^{L} m_l +m -M \Big)
\end{eqnarray}

The properties of the partition function eq.(\ref{eq.Z}), the value of the condensation threshold,
the size of the condensate and the value of $p(m)$ were examined in great detail in \cite{evans_canonical_2006}
and further extended to the presence of two constraints 
\cite{filiasi_condensation_2013,szavits-nossan_condensation_2014-1}.

In the thermodynamic limit $L,M \rightarrow +\infty$ with $M/L$ finite, three cases
are examined in \cite{evans_canonical_2006}: $f(m)$ decreases faster than the exponential,
slower than $m^{-2}$, or slower than the exponential but faster than $m^{-2}$.

For example, when $f(m)$ is a paretian distributions $1/m^{-\gamma}$ with
$\gamma \in ]2,3]$, the typical outcome is not condensed, but a phase transition 
exists, governed by the mean value $\langle f \rangle$ of $f(m)$. If $\langle f \rangle$ is smaller than the average 
density $M/L$, condensation occurs: a single site will contain the excess mass.
If $\langle f \rangle > M/L$, the system is in the fluid phase. In the limit case 
$\langle f \rangle = M/L$, the system is in the critical state. 
Condensation does not correspond to an observed phenomenon in available data, 
and will not be considered here, thus we suppose that $\langle f \rangle \geq M/L$.
Nevertheless we must keep in mind that at the critical value $\langle f \rangle = M/L$ 
or in its neighborhood, sampling algorithms can fail
because of the phase transition.

If $\gamma$ belongs to the interval $[1,2]$, $f(m)$ has a very broad tail. 
The variables $M_i$ experience large fluctuations, and the sum $\sum_{i=1}^N M_i$ has
condensed typical outcomes \cite{filiasi_condensation_2013}. 
The theoretical analysis of $Z(M, L)$ in \cite[7.2]{evans_canonical_2006} shows
that no phase transition occur, however a pseudocondensate -i.e. a bump in the tail of $p(m)$- emerges.

The expression of the marginal probability $p(m)$ is another interesting result.
While $f(m)$ is known, $p(m)$ is unknown a priori, because it results from 
incorporating the constant-sum constraint $\delta \Big( \sum_{l=1}^{L} m_l -M \Big)$ 
in the sampling process.
In \cite{evans_canonical_2006} several expressions of $p(m)$ are reviewed. 
In the grand-canonical ensemble approximation that neglects the global interaction between
the particles, $p(m)=f(m) e^{-\mu m}$,
where $\mu$ is the negative of the chemical potential.
A similar form is found in \cite[§3.1]{filiasi_condensation_2013} with the Density Functional Method.
$\mu$ must be found so that:
\begin{equation} 
\rho=M/L=\int_0^{+\infty} m f(m) e^{-\mu m} dm/\int_0^{+\infty} f(m) e^{-\mu m} dm 
\label{eq.rho_ratio_quad}
\end{equation} 
Such a value doesn't exists for all densities $M/L$ when $f(m)$ decreases slower
than the exponential but faster than $m^{-2}$, which gives rise to the aforementionned
phase transition. Let us examine the consequences in the cases covered in this article. 
If the weight function $f$ is a gamma distribution, that has an exponential tail, then 
eq.(\ref{eq.rho_ratio_quad}) always has a solution. For example if $\langle f \rangle = M/L$,
then $\mu=0$ is a solution, and $f(m)=p(m)$.

If the weight function $f$ is a lognormal distribution -i.e. with a subexponential tail-
such that $\langle f \rangle > M/L$, then eq.(\ref{eq.rho_ratio_quad}) still has a solution.

In the canonical ensemble, more detailed closed form expressions were reported in \cite{evans_canonical_2006}.

\section{Numerical experiments}
\label{sec.numeric.experiments}

In this section the methods employed and results observed in several experimental settings 
are discussed: in sec. \ref{sub.sampling}, the different flavors of the hit-and-run
sampler are compared. In sec. \ref{sub.sampling.results} the the theoretical
marginals values of wealth and income for the partially agregated BMW model 
are examined and compared to sampled values.

\subsection{Sampling methods}
\label{sub.sampling}

In this section we present briefly the algorithms used to simulate the terms
of the sum $\sum X_i = cst$, where $(X_i)$ are iid random variables, the 
distribution of which are constrained by the weight functions $f(m)$.
The parameters of functions $f(m)$ are fit to empirical data, whether income or
wealth, as seen in sec. \ref{sub.wealth.income}.

In order to compute all the variables in the economic model in sec.\ref{theory}
knowing these samples, one can simply use the different relations in the echelon form
of eq.(\ref{eq.rref}).

The Hit-and-run (HR) sampler is a standard Monte-Carlo algorithm that can be used 
in this case. If $f$ is uniform, it can be proved to converge to a uniform distribution,
with a zero rejection rate \cite{zabinsky_hit-and-run_2013}.
The HR sampler was extended to the non-uniform case, with the drawback of losing 
this last property. The issue of decreasing the rejection rate to accelerate sampling
is still an open problem \cite{shao_efficient_2013}. 
Convergence is guaranteed in $\mathcal{O}^*(d^4)$ for a large class
of target distributions in dimension $d$, 
provided that the HR implementation uses the Hypersphere Direction (HD) scheme to
update the direction during the random walk inside the polytope given by linear
constraints. Since HR is a Monte-Carlo algorithm, successive samples are correlated.
One solution to mitigate this problem is to keep only a fraction of the generated
samples, thanks to a thinning factor \cite{tervonen_hit-and-run_2013}.

Coordinate Direction (CD) is another way to update the direction,
simpler to implement because it doesn't require any change of basis. 
It can be modified to converge quicker than HD, as shown in \cite{filiasi_condensation_nodate}.
However, on the opposite of HD, its convergence is not guaranteed \cite[p.724]{zabinsky_hit-and-run_2013}.

Several works in the field of metabolic networks research use HR, to sample feasible
metabolic flows in a uniform \cite{wiback_monte_2004,  braunstein_estimating_2008}
or possibly non-uniform way \cite{capuani_quantitative_2015}.
SFC macro models were also studied with this algorithm in \cite{hazan_volume_2017}.

\subsection{Numerical results}
\label{sub.sampling.results}

\begin{figure}[htbp]
\centering
\subfigure[]{\includegraphics[width=6cm]{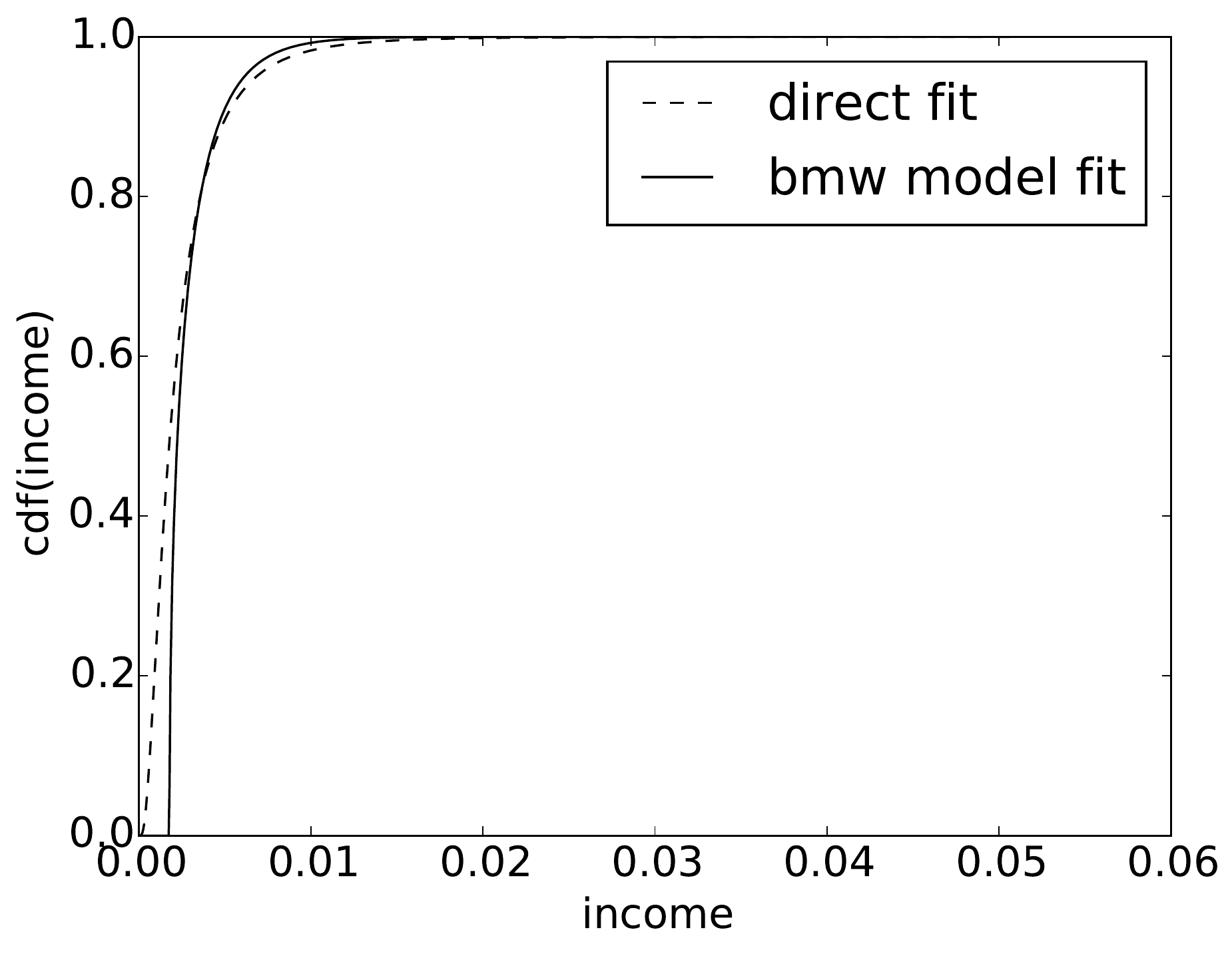}}
\subfigure[]{\includegraphics[width=6cm]{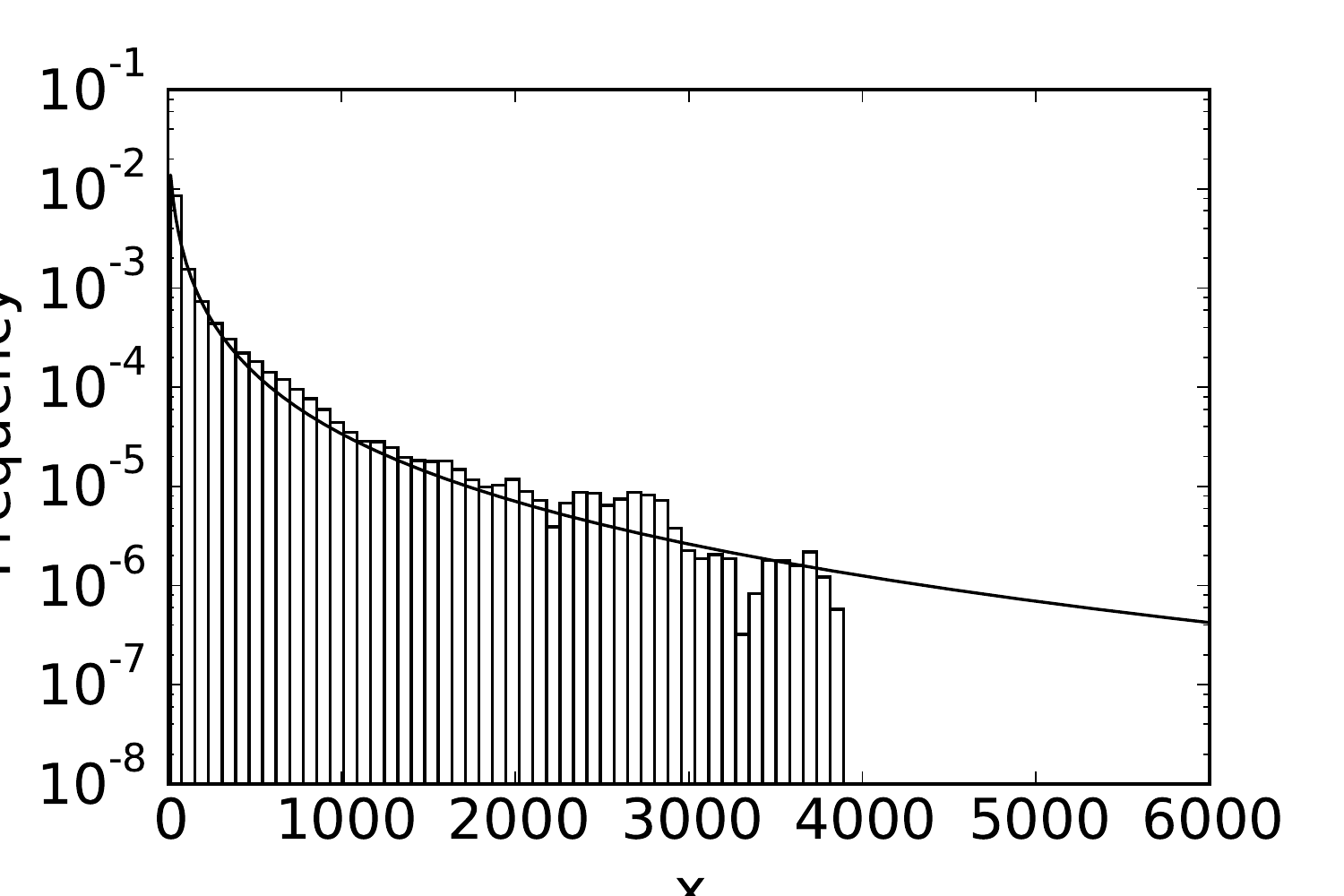}}
\caption{(a) Comparison between the direct fit of $WB_s$ to WID income data, and 
the cdf of $WB_s$ derived from the direct fit of $M_i$ to WID wealth data
(b) histogram of HR-HD wealth samples in dimension $d=100$. 
Plain line is the weight function $f$ corresponding to a lognormal distribution
with shape $\sigma=1.72$. Scaling is such that $\langle f \rangle = M/L$.
Thinning factor is 1000.
}
\label{fig.histo.sample}
\end{figure}

With the equation eq.(\ref{eq.WBs_to_M}) that links wealth to income
for each agent, one can compute the cdf of $WB_{s,i}$ from the distribution of $M_i$
that was modelled in sec. \ref{sub.wealth.income} and fit to wealth data.
The obtained cdf can be compared with the direct fit of $WB_{s,i}$ to WID income data.
If the parameters of the BMW model depicted in section \ref{theory} are unknown,
then can then be estimated, by minimizing the distance betweend the two cdf.
If they are known from previous studies, the procedure serves as a consistency check.
An illustration is shown in Fig.\ref{fig.histo.sample}(a), 
and provides the set of parameters presented in Tab. \ref{tab.bmw params}.
$r, ~\delta, ~k$ are provided by the litterature. $\alpha_0, ~\alpha_1,~\alpha_2$ are
the free parameters that need to be estimated.
Let us emphasize the necessity to rescale all the variables because of the full 
determination of $\sum M$ given the set of parameters, as seen in eq.(\ref{eq.sum_Mi}).

\begin{table}[htbp]
\centering
\begin{tabular}{cc}
\hline
Parameter & Value  \\
\hline
$\alpha_0$ & 0.001\\
$\alpha_1$ & 0.75\\
$\alpha_2$ & 0.02\\
$r$ & 0.03\\
$\delta$ & 0.1\\
$k$ & 5.5 \\
\end{tabular}
\caption{Values of parameters used in Fig. \ref{fig.histo.sample}(a).}
\label{tab.bmw params}
\end{table}

Fig.\ref{fig.histo.sample}(b) represents the histogram of wealth samples generated by a
HR-HD sampler, that approximates $p(m)$. One can notice that $p(m)$ is close to the
weight distribution $f(m)$, which is expected as mentionned in sec. \ref{sub.sampling},
except in the region of the tail, because of the finiteness of the sample.

As was recalled in sec. \ref{sub.sampling},  there is no general proof of convergence for 
HR-CD, that would be available for a large class of distributions. We found some
examples where HR-CD failed to sample correctly
the distribution of weights $f(m)$ with very broad tails, even though we stayed 
in the typical fluctuation regime $\langle f \rangle \geq M/L$:
first a monotone polynomial fit $m \rightarrow \frac{1}{m}\sum_{i=0}^N \beta_i \log(m)^i$, used to model the wealth;
then a lognormal distribution with shape parameter $\sigma=1.72$. 
This failure is not mitigated using the acceleration method in \cite{filiasi_condensation_nodate}.
It calls for some improvement on the algorithms currently used (HR-HD), because of the
high rejection rate due to sampling a high-dimensional distribution.


\section{Discussion}
\label{sec.discussion}

In this paper a partially aggregated SFC model was studied. This is a restriction
to more general problems, such as disaggregated models, and also random
topology models, where the connectivity is not fixed.
The partially aggregated model is useful as a limit case, because some additive results
can be obtained, as was seen above.
However it adds restriction on the type of distribution considered. For example,
one may want to decouple the distributions of wealth and income. This can still be 
achieved with the partially aggregated model, for example considering $\alpha_{0,i}$ as
random variables, rather than fixed constants.

Furthermore, one may question the usefulness of sampling as a tool to study the marginal
distributions of variables in the model, since closed form expressions of the cdf can
be obtained. Once again, the partially aggregated model is a limit case. In the 
case of fully disaggregated models, it is not likely that a simple equation will
relate $WB_{s,i}$ and $M_i$. Some constant quantities might appear though, as in the case of
metabolic networks \cite{de_martino_identifying_2014}.


\section{Conclusion}
\label{sec.conclusion}

In this article the problem of finding the marginal probabilities of the variables in
a steady-state partially aggregated SFC macroeconomic model is addressed, with
distributional constraints imposed on individual wealth variables.
This last feature can be thought of as expressing the fact that the distribution
of wealth or income are exogenous, as the result of an evolving balance of power between
capital and workforce.
This improves over \cite{hazan_volume_2017} where an SFC model was seen as a
Constraint Satisfaction Problem, but without \emph{a priori} constraint on distributions.

The following results are reported: 
\begin{itemize}
\item using standard algebra, we find that the linear system of equations that 
forms part of the problem is underdetermined and amounts to a constant sum equation. 
\item we recall relevant results from the theory of mass transport, that give an approximation
of the marginal probabilities, given the weight distributions. Since condensation is not
a admissible outcome for the system state, we deduce some constraint on the parameters of the weight distribution.
\item we fit empirical wealth data to a classical distribution, and compute the 
corresponding cdf for income. The latter is compared to an direct empirical fit of 
income data, which allows us to estimate some free parameters.
\item finally, sampled solutions to the initial problem are shown, with various 
hit-and-run algorithms. 
\end{itemize}

This last result can be used by practitionners in the SFC community that are 
interested in distributional phenomena, and can be compared with the results
obtained using time-averaged SFC/ABM models.

In future work we will compare the limit results obtained here in the case of a
partially aggregate model to more general settings. 
First an extension to disaggregated models will be examined. Thanks to recent
developments in the analysis of metabolic networks 
we hope to overcome the curse of dimensionality;
which will be compared to accelerations of hit-and-run algorithms such as \cite{shao_efficient_2013}.
An extension to random network, following \cite{martino_von_2007, bianconi_flux_2008, seganti_boolean_2013},
will also be addressed.

Furthemore in this article, we limited our scope to BMW models, that form just a part 
of SFC models. More models, that fall in the category of linear dynamical systems, 
will be analyzed, and more empirical data will be included.

Lastly, the dynamical behavior may be of interest, for example steadily growing 
economies, or stability properties \cite{anand_stability_2009}, that are related to the occurrence of crises.

\appendix

\section{Transaction matrix for the BMW model}

Tab. \ref{bmw.transactions} sums up the different transactions in the modelled economy.
Firms and banks are each represented by a single agent. The number of households is set to $nw=3$
in this example but can take any strictly positive value.

\begin{table}[htbp]
\centering
	\begin{tabular}{lllllllll} %
	\hline		
 	& \multicolumn{3}{c}{Households}  & \multicolumn{2}{c}{Production Firms} & \multicolumn{2}{c}{Banks} & $\sum$ \\
 	\cmidrule(lr{.75em}){2-4}  \cmidrule(lr{.75em}){5-6} \cmidrule(lr{.75em}){7-8}
	&   &   &   & \multicolumn{1}{c}{Current} & \multicolumn{1}{c}{Capital}  & \multicolumn{1}{c}{Current} & \multicolumn{1}{c}{Capital}  & \\
	& 1 & 2 & 3 & 1 & 1 & 1 & 1 & \\
	\hline
	Consumption & -$C_{d1}$ &  &  & $C_{s1}$ &   &  &  & 0\\
	            &  & -$C_{d2}$ &  & $C_{s2}$ &   &  &  & 0\\
	            &  &  & -$C_{d3}$ & $C_{s3}$ &   &  &  & 0\\
	\hline            
	Investment &  &  &  &  $I_{s1}$ & -$I_{d1}$ &  &  & 0\\            
	           &  &  &  & $I_{s2}$ & -$I_{d2}$  &  &  & 0\\            
	\hline            	
	Wage & $WB_{s1}$ &  &  & -$WB_{d1}$   &  &  &  & 0\\
	     &  & $WB_{s2}$ &  & -$WB_{d2}$   &  &  &  & 0\\
	     &  &  & $WB_{s3}$ & -$WB_{d3}$   &  &  &  & 0\\
	\hline            	
	Depreciation &  &  &  & -$AF_1$ & $AF_1$ &  &  & 0\\            
	\hline            	
	Interest on loans &  &  &  & -$IL_1$ & &  $IL_1$  &&  0\\            
	\hline            	
	Interest on deposits & $ID_1$ &  &   &  & &  -$ID_1$ &  & 0\\            
					     &  & $ID_2$ &   &&  & -$ID_2$ &  &   0\\            
						 &  &  & $ID_3$   &  &  & -$ID_3$ &   & 0\\            	
	\hline
	Change in loans &  &  &  &  $\Delta L_1$ &  &   & -$\Delta L_1$  & 0\\            
	\hline
	Change in deposits & -$\Delta M_1$ &  &  &  &  &  & $\Delta M_1$ & 0\\            
	                   &  & -$\Delta M_2$ &  &  &  &     &  $\Delta M_2$   & 0\\            
	                   &  &  & -$\Delta M_3$ &  &  &  &   $\Delta M_3$ & 0\\            	
	\hline
	$\sum$ & 0 & 0 & 0 & 0 & 0 & 0  & 0 & 0\\            
	\hline
	\end{tabular}
\caption{\label{bmw.transactions}Transaction matrix of the BMW model 
with many households, on firm and one bank: agents $nw=3$, $nf=1$, $nb=1$. 
Households must buy goods from the only available firm.
The firm is buying capital goods from itself.
In the stationay case, $\Delta L=\Delta M=0$.}
\end{table}

\section{Econometric data}
\label{appendix.data}

Econometric data in sec. \ref{sub.wealth.income} are taken from the World Inequality Database 
\url{http://wid.world}. Tab. \ref{tab.WID} extracts some corresponding information among
the documentation supplied by the authors.
For both variables ``the base unit is the individual (rather than the household) but resources are split equally within couples.  The population is comprised of individuals over age 20''.

\begin{table}[htbp]
\centering
	\begin{tabular}{llp{7cm}lll} %
	\hline
	Variable Name & Year & Description & Unit & Ref. & Code \\
	\hline
	Net personal wealth & 2012 &Net personal wealth threshold value at a given percentile.  Net personal wealth is  the total value of non-financial and financial assets (housing, land, deposits, bonds, equities, etc.) held by households, minus their debts.  & EUR constant 2015 & \cite{garbinti_income_2016} & thweal992j \\
	Fiscal income  & 2012&  Fiscal income threshold value at a given percentile.  Fiscal income is defined as the sum of all income items reported on income tax returns, before any deduction. It includes labour income, capital income and mixed income.  & EUR constant 2015 & \cite{garbinti_income_2016} & tfiinc992j
	\end{tabular}
\caption{\label{tab.WID} Extract of the WID documentation.}
\end{table} 
 
\section{Acknowledgements}

Open-source software were used to perform this research: Python, Cython, R, Sage, \LaTeX.


\bibliographystyle{elsarticle-num} 
\bibliography{HRF+eco}





\end{document}